\documentclass[prb,preprint,a4paper,floatfix]{revtex4-1}
\usepackage{amsmath}
\usepackage{fancybox}
\usepackage{eepic}
\usepackage{times}
\usepackage{latexsym}
\usepackage{pifont}
\usepackage{epsf}
\usepackage{upgreek}
\usepackage{epsf}
\usepackage{amssymb}
\usepackage{graphicx}
\usepackage{bm}

\def\jpcm{J. Phys.: Cond. Matt.}   

\def\prb{Phys. Rev. B}

\def\prl{Phys. Rev. Lett.}

\begin{document}
\title{Nonconservative dynamics in long atomic wires}
\author{Brian Cunningham}
\address{Atomistic Simulation Centre, School of Mathematics and Physics, 
        Queen's University Belfast, 
        Belfast BT7 1NN, UK} 
\author{Tchavdar N Todorov}
\address{Atomistic Simulation Centre, School of Mathematics and Physics, 
        Queen's University Belfast, 
        Belfast BT7 1NN, UK} 
\author{Daniel Dundas}
\address{Atomistic Simulation Centre, School of Mathematics and Physics, 
        Queen's University Belfast, 
        Belfast BT7 1NN, UK} 
\begin{abstract}
The effect of nonconservative current-induced forces on the ions in a defect-free metallic nanowire is investigated using both steady-state calculations and dynamical simulations.  Non-conservative forces were found to have a major influence on the ion dynamics in these systems, but their role in increasing the kinetic energy of the ions decreases with increasing system length.  The results illustrate the importance of nonconservative effects in short nanowires and the scaling of these effects with system size.  The dependence on bias and ion mass can be understood with the help of a simple pen and paper model.  This material highlights the benefit of simple preliminary steady-state calculations in anticipating aspects of brute-force dynamical simulations, and provides rule of thumb criteria for the design of stable quantum wires.
\end{abstract}
\maketitle

%
%
%
\section{Introduction} 
The miniaturisation of electronic devices results in increasing current densities.  These current densities generate large forces on individual atoms with considerable effects on the functionality and stability of the device.  Understanding the mechanisms by which electrons and ions in a nano-conductor exchange energy is therefore essential.

The current-induced force on an atom consists of the average force and fluctuating forces.  Fluctuating forces are due to the corpuscular nature of electrons and are responsible for processes such as Joule heating \cite{horsfield:2004,galperin:2007,lu:2012}.  The average force on the other hand contains, among other contributions, the familiar electron wind force \cite{landauer:1974,sorbello:1997,todorov:philmagb:2000,diventra:2002,brandbyge:2003,zhang:2011}.  In recent years the wind force has become the focus of renewed attention due to the realisation that it is nonconservative \cite{sorbello:1997,stamenova:2005,dundas:2009,lu:2010,todorov:2011,bode:2011,lu:2012,bode:2012,dundas:2012}.  The importance of these nonconservative forces cannot be overestimated.  Such forces can act either constructively or destructively on a nanoscale device.  Constructive work leads to the possibility of nanoscale engines, while destructive work can act as an activation mechanism for electromigration and device failure \cite{taychatanapat:2007}.  Non-conservative forces may be a prime candidate for explaining apparent heating in atomic wires \cite{Tsutsui:2008,Tsutsui:2007} far above that expected from Joule heating alone \cite{todorov:2001, smit:2004}.

It is shown in \cite{dundas:2009, lu:2010} that nonconservative effects in atomic wires require near degenerate vibrational mode frequencies.  Current can couple such modes to produce new modes that grow or decay in time.  In the simplest case of two modes, the new modes are abstract rotors of opposite angular momentum \cite{todorov:2011}, one of which is driven by the current and the other is attenuated.  We will refer to growing or decaying modes, generically, as waterwheel modes.  The likelihood of the formation of waterwheel modes should, in general, increase with the number of near degeneracies.  Defect-free metallic nanowires are of special interest for these effects.  The reason is that the symmetric part of the current-induced contribution to the dynamical response matrix \cite{dundas:2012} vanishes to lowest order in the bias.  This symmetric part is controlled by the real part of the electronic density matrix in the real space representation.  In a perfect wire, left and right travelling electronic wave functions come in complex conjugate pairs and hence the repopulation of these states under small bias leaves the real part of the electronic density matrix unchanged.  

This eliminates a central impediment to nonconservative dynamics, namely bias-induced frequency renormalisation, which lifts the degeneracies discussed above and competes with nonconservative energy build-up.  These considerations make metallic nanowires a prime candidate for the observation of nonconservative effects on a grand scale.  

In this paper we investigate nonconservative effects in long defect-free 1-D atomic wires.  As a result of the competition between nonconservative forces and the electronic friction the ionic kinetic energies saturate at a bias-dependent steady state.  The kinetic energy per atom (and hence effective steady-state temperature) decreases with increasing wire length and increases with atomic mass, while (for long chains) the saturation current is determined solely by the atomic mass.  The results are compatible with a simple pen and paper model and furnish criteria in the design of stable atomic scale leads.  
%
%
\section{Methods}

We employ two methods: static steady-state transport calculations, and nonequilibrium nonadiabatic electron-ion molecular dynamics in the Ehrenfest approximation with electronic open boundaries \cite{dundas:2009}.  
In both cases the electronic structure is described in a spin-degenerate nearest-neighbour single-orbital orthogonal tight-binding model \cite{sutton:2001} with noninteracting electrons.  The hopping integral between sites $m$ and $n$ is 
\begin{equation}\label{hamiltonian}
H_{mn}=-\displaystyle\frac{\epsilon c}{2}\left(\displaystyle\frac{a}{R_{mn}}\right)^{q},
\end{equation} 
where $R_{mn}$ is the separation between the sites.  The on-site energies are set equal to zero.  The pair potential between sites $m$ and $n$ is 
\begin{equation}\label{pair}
P_{mn}=\epsilon\left(\displaystyle\frac{a}{R_{mn}}\right)^p.
\end{equation}
The tight-binding parameters are those for gold \cite{sutton:2001}: $a=4.08~{\rm \AA}$ is a length scale; $\epsilon=0.007868~{\rm eV}$ is an energy scale; $c=139.07$ is a dimensionless constant controlling the relative contributions of electronic binding and the repulsive pair potentials; $q=4$ and $p=11$ are the inverse power exponents.  We set the lattice parameter to $2.37{\rm \AA}$, below the equilibrium value of $2.52{\rm \AA}$, to suppress a Peierls transition and the resultant band gap that tend to occur during relaxation otherwise.  The hopping integral, $H$, then is $-4.78~{\rm eV}$.  The hopping integral and pair potential are truncated between first and second neighbours by a smooth tail. 
%
%
\subsection{Static current-carrying steady state}

\begin{figure}[t!]
\includegraphics[trim=0.0cm 0cm 0cm 0.0cm, clip=true,width=0.72\textwidth, totalheight=0.195\textheight, angle=0]{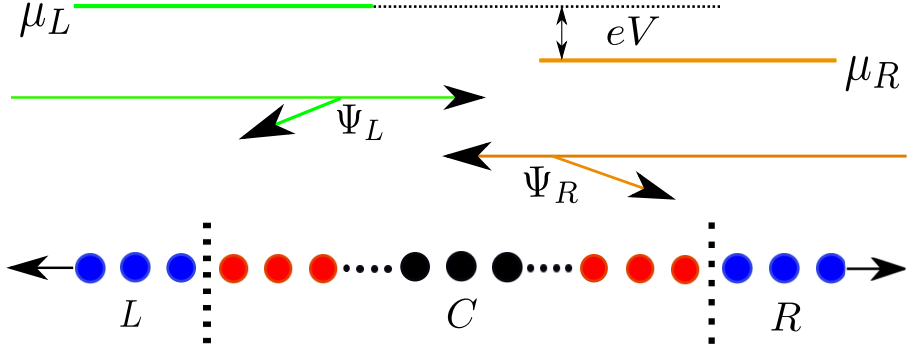}
\caption{Device {\it{C}} is connected to semi-infinite electrodes $L$ and $R$ (blue).  $C$ consists of a central sub-region (black) of varying lengths, which later will be treated dynamically, while holding the red sub-regions rigid. In the static Landauer picture, left- and right-travelling Lippmann-Schwinger scattering electron wave functions $\{\Psi_L\}$ and $\{\Psi_R\}$  are populated with Fermi-Dirac distributions $f_L$ and $f_R$, corresponding to electrochemical potentials $\mu_{L,R}=\mu\pm eV/2$, where $\mu$ is the equilibrium chemical potential.  The electrochemical potential difference $eV=\mu_L-\mu_R$ generates a net flux of electrons.}
\label{1d-chain}
\end{figure}

The static approach employs the Landauer picture, figure \ref{1d-chain}.
The 1-electron steady-state density matrix is
\begin{equation}\label{densmat}
\hat{\rho}({ V},{\bm R}) = \int_{-\infty}^{+\infty}\hspace{-0.0cm}\left[f_L(E)\hat{D}_L(E)+f_R(E)\hat{D}_R(E)\right]{\rm d}E,
\end{equation}
where $\hat{D}_{i}(E)$, with $i=L,R$, is the density of states operator for the scattering states $\{\Psi_{i}\}$ with occupations $\{f_{i}(E)\}$.  The total density of states operator, $\hat{D}(E)=\hat{D}_L(E)+\hat{D}_R(E)$, can be expressed in terms of the retarded and advanced Green's functions: $\hat{D}(E) =[\hat{G}^{-}(E)-\hat{G}^{+}(E)]/{2\pi{\rm i}}$.  Spin degeneracy is subsumed into $\hat{D}_i$.

The force on ionic degree of freedom $\nu$ due to electrons is 
\begin{equation}\label{force}
F_{\nu}(V,{\bm R}) = {\rm Tr}\left\{\hat{\rho}(V,{\bm R})\hat{F}_{\nu}({\bm R})\right\}.
\end{equation}
$\hat{F}_{\nu}({\bm R})=-\partial \hat{H}({\bm R})/\partial R_\nu$, where $\hat{H}({\bm R})$ is the electronic Hamiltonian as a function of the ion coordinates, ${\bm R}$.  In general, $\nu$ labels an atom and direction; in the present case we only have longitudinal displacements and $\nu$ labels just the atom.  

Small-amplitude atomic motion about a reference geometry, ${\bm R}$, is characterised by the steady-state dynamical response matrix
\begin{equation}\label{dynresp}
K_{\nu\nu'}({ V},{\bm R})=-\displaystyle\frac{\partial F_{\nu}({V},{\bm R})}{\partial R_{\nu'}}+\frac{\partial^2P({\bm R})}{\partial R_{\nu'}\partial R_{\nu}},\vspace{0.1cm}
\end{equation}
where $P({\bm R})$ is the sum of pair potentials from (\ref{pair}).  This matrix can further be split into an equilibrium part and a current-induced correction $\Delta K_{\nu\nu'}({V},{\bm R})$.  
$\Delta K_{\nu\nu'}$ can then be decomposed into a symmetric and an antisymmetric part, $\Delta K_{\nu\nu'}=S_{\nu\nu'}+A_{\nu\nu'}$ \cite{dundas:2012}, where  
\begin{align}
S_{\nu\nu'}({ V},{\bm R})
=&\sum_{i=L,R}\int_{\mu}^{\mu_i}\hspace{-0.2cm}\Big(2~{\rm Re}~{\rm Tr}\left\{\hat{F}_{\nu}\hat{R}(E)\hat{F}_{\nu'}\hat{D}_i(E)\right\}\nonumber \\
&-{\rm Tr}\left\{\frac{\partial\hat{F}_{\nu}}{\partial R_{\nu'}}\hat{D}_i(E)\right\}\Big) {\rm d}E, \\[0.2cm]
\label{curl_as}
A_{\nu\nu'}({ V},{\bm R})
=&2\pi\displaystyle\sum_{i=L,R}\int_{\mu}^{\mu_i}\hspace{-0.3cm}{\rm Im}~{\rm Tr}\left\{\hat{F}_{\nu}\hat{D}(E)\hat{F}_{\nu'}\hat{D}_i(E)\right\}{\rm d}E,
\end{align} 
with $\hat{R}(E)=\{\hat{G}^{-}(E)+\hat{G}^{+}(E)\}/2$.  All quantities inside the traces above are themselves functions of ${\bm R}$.  The antisymmetric part in (\ref{curl_as}) is the origin of the nonconservative forces \cite{dundas:2012}.  It makes the dynamical response matrix non-Hermitian with the possibility of complex frequencies describing motion that grows or decays exponentially in time.  
The larger the anti-symmetric part the greater the possibility of these nonconservative effects.  
%
%
\subsection{Dynamical transport simulations}

What do complex mode frequencies imply physically?  Will the kinetic energy of the ions increase indefinitely leading to the eventual rupture of the wire?  In a real wire we have the cooling effect of the electronic friction, further velocity-dependent forces \cite{lu:2010}, and possibly large and violent departures from the perfect wire geometry.  
We address this complexity by direct nonequilibrium nonadiabatic molecular dynamics simulations, within the Ehrenfest approximation, using the tight-binding model above \cite{dundas:2009}.  Current is generated by the open-boundary method of \cite{mceniry:2007}, with $S$ an 800-atom long 1-D chain and $C$ consisting of the 300 central atoms, a subset of which (black in figure~\ref{1d-chain}) are treated dynamically.  The electrodes are 250 atoms each, and the sink and source terms are applied to all electrode atoms with $\Gamma=0.5~{\rm eV}$ and $\Delta=0.0005~{\rm eV}$.  The dynamical simulations employ the Ehrenfest approximation, which treats the nuclei as classical particles interacting with the mean instantaneous electron density.   This approximation suppresses correlations between electronic and ionic fluctuations and the microscopic noise in the force exerted by the electrons on the ions.  This in turn suppresses Joule heating.  This crucial limitation of Ehrenfest dynamics, however, will work to our advantage: it leaves nonconservative current-induced forces as the only energy injection mechanism into the atomic motion, enabling us to isolate and study its effect.  In addition, as we will see later, Joule heating would only have a weak effect in the long-time dynamical regime reached by the system.  The additional cooling effect of lattice conduction out of the mobile region is also not incorporated, to give us an upper bound on what the nonconservative forces can do.
%
%
\section{Results and Discussion}
\subsection{Preliminary static calculations}\label{res_st}

We examine the mode frequencies in a defect-free atomic wire as a function of the number of mobile atoms, $N$.  The mode frequencies are determined from the square root of the eigenvalues of the dynamical response matrix, equation (\ref{dynresp}), for relaxed mobile atoms (nearby geometries produce qualitatively similar phonon structure). 

\begin{figure}[t!]
\includegraphics[trim=0.0cm 0cm 0cm 0cm, clip=true, width=0.75\textwidth, totalheight=0.4\textheight, angle=0]{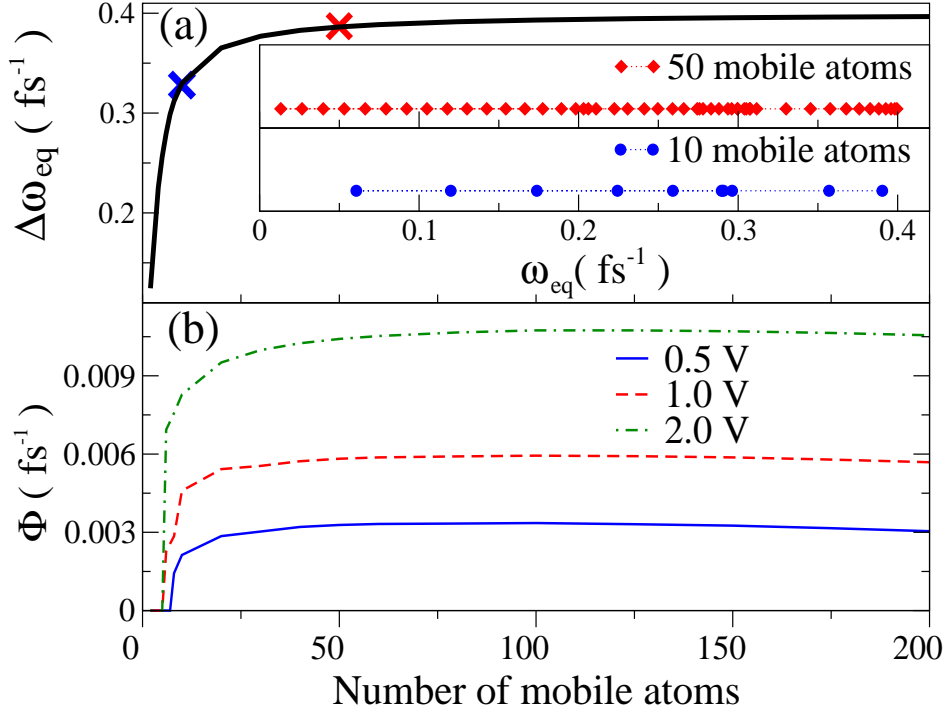}
\caption{(a): Range of equilibrium mode frequencies as a function of mobile region length, $N$, for atoms with mass 10 a.m.u. (the inset displays the individual frequencies for the two lengths marked with an $\times$).  (b):  $\Phi$, equation~(\ref{AMMFIP}), as a function of $N$ for three biases.}
\label{eqm_freq}
\end{figure}

Figure \ref{eqm_freq}(a) shows the range of equilibrium frequencies,  
$\Delta \omega_{\rm eq}={\rm max}(\omega_{\rm eq}) - {\rm min}(\omega_{\rm eq})$,
as a function of $N$.  The range in figure \ref{eqm_freq}(a) saturates with $N$ at the phonon bandwidth.  Thus, the typical spacing between frequencies decreases and, in longer systems, more waterwheel modes should form under bias. 

\begin{figure}[t!]
\includegraphics[trim=0cm 0cm 0.0cm 0.0cm, clip=true, width=0.75\textwidth, totalheight=0.4\textheight, angle=0]{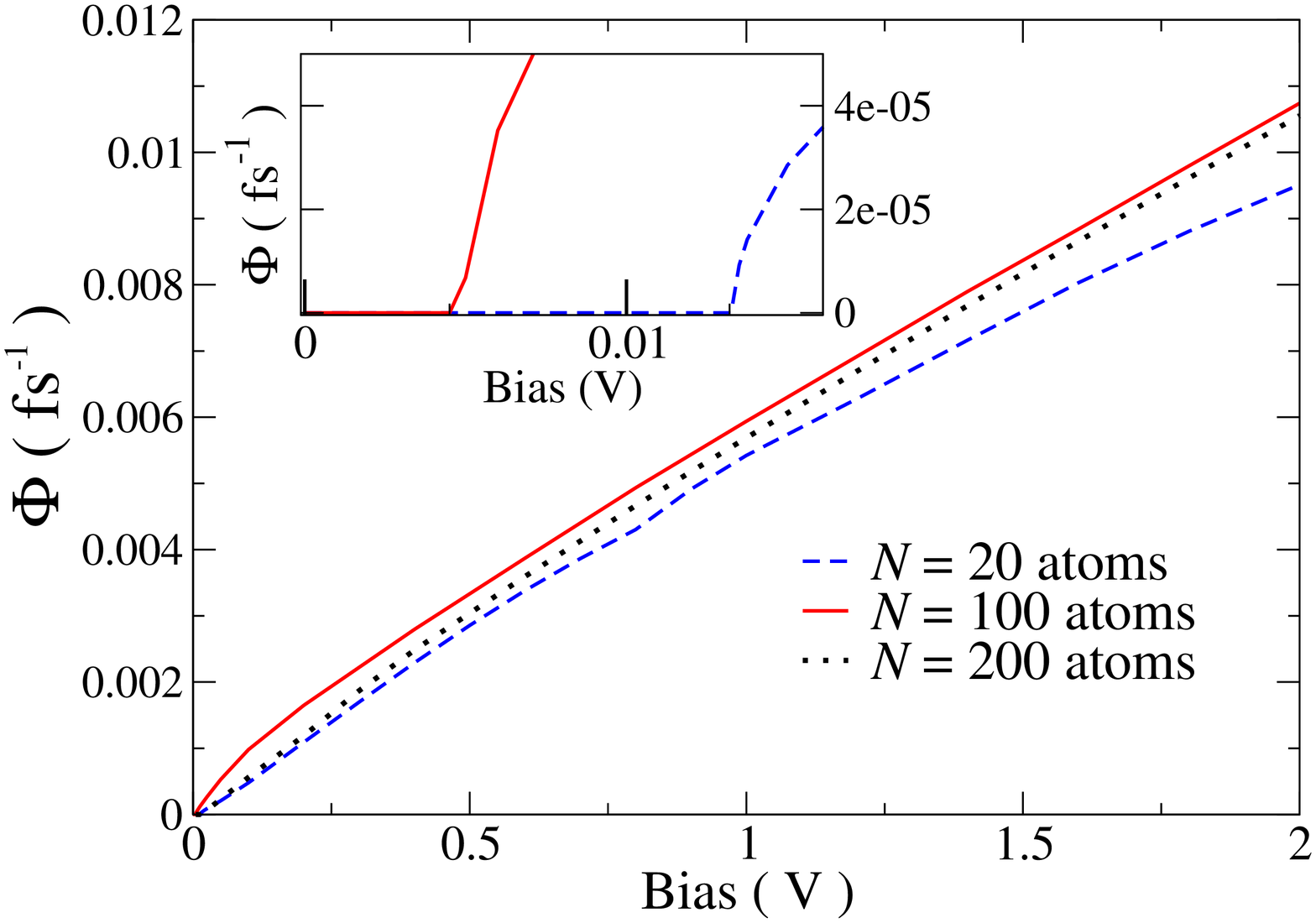}
\caption{$\Phi$, equation~(\ref{AMMFIP}), as a function of applied bias for different device lengths.  Since the number of near degenerate modes increases with wire length, figure~\ref{eqm_freq}(a), the critical bias required to overcome frequency mismatches should decrease with wire length and this can be seen in the inset.  The atomic mass is 10 a.m.u..}
\label{sum_vs_bias}
\end{figure}

Next we calculate the mode frequencies under bias.  Since the dynamical response matrix is now nonHermitian, complex eigenvalues are possible and appear in complex conjugate pairs.  Mode frequencies also come in conjugate pairs, corresponding to growing or decaying waterwheel modes.  For a given bias, the number of such pairs increases in a staircase-like fashion with the number of mobile atoms.  Next we form the quantity\begin{equation}\label{AMMFIP}
\Phi=\frac{1}{N}\sum_{\alpha=1}^N|{\rm Im}\left(\omega_{\alpha}\right)|,
\end{equation}
as a function of $N$ and $V$ in figures \ref{eqm_freq}(b) and \ref{sum_vs_bias}.
To within a proportionality constant, this quantity provides a notional measure of the rate of work, per atom, due to nonconservative forces.  We see that beyond $N\gtrsim 40$, $\Phi$ saturates with mobile region length and increases linearly with bias.  For a given $N$ and $V$, the modes with appreciable imaginary parts to their frequencies tend to be a small fraction (which increases with bias) of the total number of modes.  Both the imaginary and real parts of their frequencies are closely clustered together.  Physically, these modes correspond to the directional stimulated emission, or absorption, of travelling phonons \cite{todorov:2011}. These findings suggest that nonconservative current-induced dynamics in longer wires might exhibit certain bias-dependent, length-independent characteristics.  This is now investigated by full dynamical simulations in which the nonconservative forces compete with the electronic friction. 
%
%
%
\subsection{Current-driven dynamics}\label{t-dsims}

\begin{figure}[t!]
\includegraphics[trim=0cm 0cm 0cm 0cm, clip=true, width=0.72\textwidth, totalheight=0.36\textheight, angle=0]{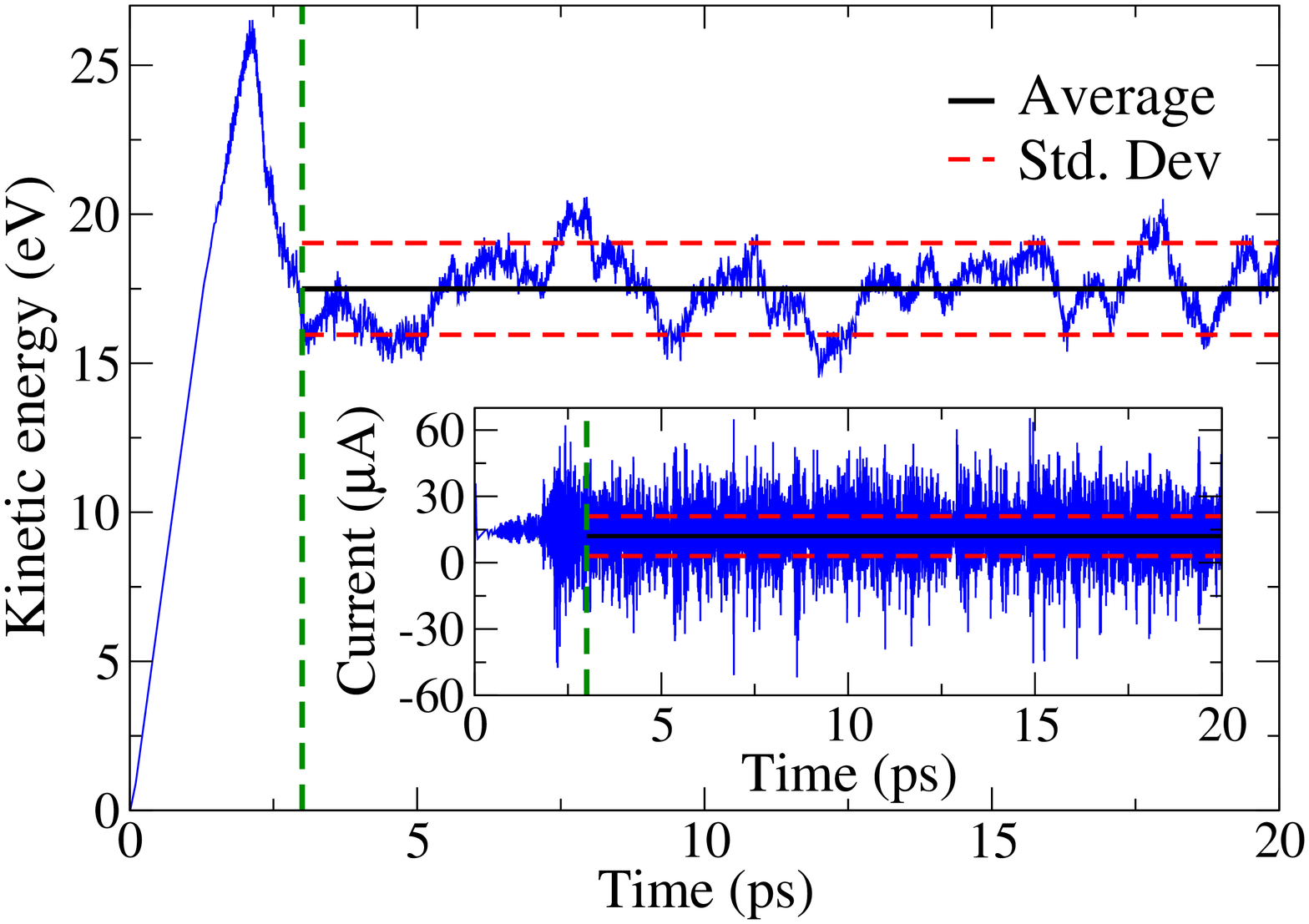}
\caption{Combined kinetic energies of all 200 mobile atoms with mass 10~a.m.u. as a function of time for a bias of 0.5~V with the inset displaying the bond current for the bond in the middle of the chain as a function of time.  The quantities saturate after about 3 ps (vertical dashed lines), enabling us to determine average values for the energy and current.}  \label{kevstime}
\end{figure}
The dynamical simulations under bias start from the above relaxed geometry.  Figure~\ref{kevstime} shows the total kinetic energy of ions with mass 10~a.m.u. as a function of time for a device containing 200 moving atoms under a bias of 0.5~V.  The ``heating" of the ions by the nonconservative current-induced forces gives rise to the sharp initial increase in the kinetic energy.  The electronic friction, which effectively cools the ions, then kicks in.  The balance between the two causes the kinetic energy to saturate and fluctuate about a mean value.  In figure~\ref{kevstime} this happens after about 3~ps with a time-averaged total kinetic energy thereafter of $17.5\pm 1.5$~eV.  The inset in figure~\ref{kevstime} displays the bond current as a function of time for the middle bond in the chain (the bond current is a quantity that arises with atomic-orbital basis sets \cite{mceniry:2007}, and with the present tight-binding model, the bond current gives the physical current flowing between the respective two sites).
\begin{figure}[t!]
\includegraphics[trim=0cm 0cm 0cm 0cm, clip=true, width=0.675\textwidth, totalheight=0.21\textheight, angle=0]{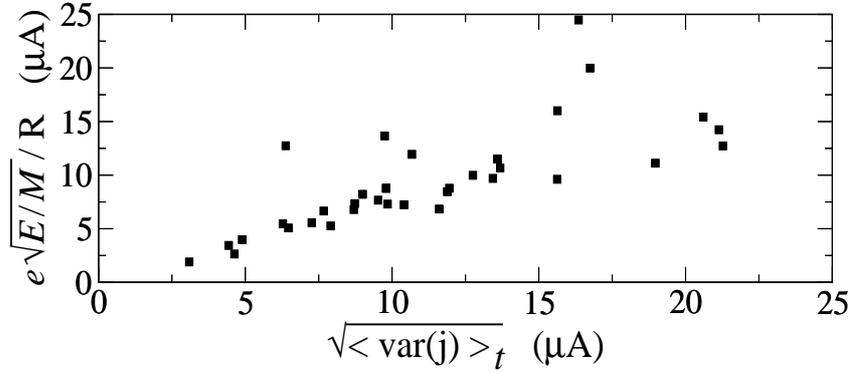}
\caption{Data for the quantities in equation~(\ref{eq_current_density}) (expressed as electric current) from a range of dynamical simulations spanning biases between 0.5~V and 2~V, masses between 1~a.m.u. and 40~a.m.u., and lengths between 10 and 200 atoms.  The spatial variance of the current is found from the middle half of the dynamical region in every snapshot and this is then time-averaged.}\label{current_vs_pos}
\end{figure}

A notable feature is the current noise in figure~\ref{kevstime}.  We expect variations in the current, even under ideal steady-state conditions, as the atomic geometry varies in time.  However, a significant contribution to the current fluctuations in the simulations comes from departures from steady-state behaviour.  They arise due to multiple dynamical electron scattering in the vibrating region and result in {\it spatial} variations of the current along the wire at any one time. These nonadiabatic current fluctuations allow a simple analytical model.  Atomic vibrations result in variations in the hopping integrals, in space and in time.  This in turn results in variations in the bottom of the electronic conduction band and thus in effective local driving fields.  We model the resultant electron dynamics with the semi-classical driven diffusion equation for the electron density $\rho$
\begin{equation}\label{eq_current}
\frac{\partial\rho}{\partial t}=D\frac{\partial^2\rho}{\partial x^2}-\sigma  \frac{\partial F}{\partial x},
\end{equation}
where $D\sim vl_{\rm tr}$, with $v$ the Fermi velocity and $l_{\rm tr}$ the electron transport mean free path, is the diffusion coefficient and $\sigma\sim Dd$ is the conductivity, with $d$ the Fermi local density of states.  $F=F(x,t)$ is the driving force field due to the breathing of the band edge caused by the motion of the ions.  To keep the model simple, we treat the phonons as dispersionless jellium phonons with a displacement field $X(x,t)$.  Then $F\sim -2H'\partial X(x,t)/\partial x$, where $H'$ is the derivative of the hopping integral with bond length\footnote{Variations in the hopping integrals result in variations in bandwidth, as opposed to rigid shifts of the band as a whole.  The resultant effective fields are different from ordinary fields.  For example, the dispersion relation for an electron in a 1-D nearest-neighbour single-orbital orthogonal tight-binding model is $E=2H\cos{\phi}$, where $E$ is the electron energy and $\phi$ is a dimensionless crystal momentum.  Therefore if $H$ varies with position, to conserve energy $\phi$ must vary as the electron propagates.  In this sense, the electron is experiencing a field.  However, the effect depends on electron energy: it vanishes in the centre of the band, where $\phi=\pi/2$ and $E=0$ irrespective of $H$, and becomes prominent near the band edges.  The simple model in equation~(\ref{eq_current}) does not account for energy-dependent effective fields.  Therefore, we must consider instead a fictitious electron-phonon coupling in which vibrations couple directly to the local electron potential (such as the tight-binding on-site energies), generating ordinary effective fields.  This conversion is {\it ad hoc}, both in amplitude and in phase, and its strength may depend on phonon wavevector.  The given $F$, with $H'$ playing the role of a coupling parameter, attempts to do that with the aim of capturing the magnitude of the current fluctuations rather than their microscopic detail.}.  Next, expand the displacement field in normal modes, $X=\sum_{k}A_k\sin{kx}\sin{\omega_k t}$.  The resultant particular integral to equation~(\ref{eq_current}) is $\Delta\rho=\sum_kB_k\sin{kx}\sin{(w_kt+\phi_k)}$, where  $B_k=2\sigma H'k^2A_k/\sqrt{w_k^2+D^2k^4}$ (the exponentially decaying transients are subsumed into the complimentary function).  From the continuity equation, $\partial \Delta\rho/\partial t+\partial \Delta j/\partial x=0$, for the fluctuating part of the particle current, we obtain $\Delta j=c\sum_kB_k\cos{kx}\cos{(\omega_k t+\phi_k)}+j_0(t)$, where $j_0(t)$ is a divergence-free part and $c=\omega /k$.  Next, we consider the spatial variance of the current: $(1/L)\int_{0}^L\Delta j^2(x,t){\rm d}x-j_0^2(t)$, where $L$ is the length of the system.  The time-average of this spatial variance then becomes
\begin{equation}\label{eq_var}
\langle{\rm var}(j)\rangle_t=\frac{1}{\pi}\int_{k_{\rm min}}^{k_{\rm max}}\frac{H'^2\sigma^2}{1+(Dk/c)^2}\frac{8ER}{Mc^2N}{\rm d}k,
\end{equation} 
where $E$ is the total ionic kinetic energy, $R$ is the lattice parameter and $M$ is the atomic mass and we have assumed equipartitioning of energy between the different modes.  From then on different regimes are possible depending on the value of $Dk/c$ at the limits of integration.  With $k_{\rm max}\sim\pi/R$, $Dk_{\rm max}/c\sim vl_{\rm tr}\pi/Rc\gg 1$, under physical conditions.  Thus we can take the upper limit to $\infty$.  But with  $k_{\rm min}=\pi/L$, $Dk_{\rm min}/c\sim vl_{\rm tr}\pi/cL$.  The average current in figure~\ref{kevstime}, under the given bias, corresponds to a transmission probability of about 0.3.  The rest of our simulations will also be characterised by transmission probabilities of that order of magnitude.  Therefore we are in a regime where $l_{\rm tr}/L$ is less than unity but not much less than unity, while $v/c\gg 1$.  In this intermediate regime, therefore, we must treat $Dk_{\rm min}/c$ as a number considerably in excess of 1.  Then equation~(\ref{eq_var}) gives 
\begin{equation}\label{eq_current_density}
\sqrt{\langle{\rm var}(j)\rangle_t}\approx \sqrt{8H'^2R^2d^2E/\pi^2M}=\frac{\zeta}{R}\displaystyle\sqrt{\displaystyle\frac{E}{M}}.
\end{equation} 
where $\zeta$ is a factor of order unity for typical parameters.  

Figure~\ref{current_vs_pos} compares the simple result in equation (\ref{eq_current_density}) against data from the whole pool of simulations that we have performed.  In figure~\ref{current_vs_pos} we calculate the spatial variance in bond current for all bonds in the middle half of the dynamical region at regular time intervals, time average these and compare with the quantity on the r.h.s. of equation~(\ref{eq_current_density}), with $E$ determined from the simulations, as in figure~\ref{kevstime}.  We see clear qualitative agreement in figure~\ref{current_vs_pos}.  This interesting dynamical current noise not only explains the current fluctuations in figure~\ref{kevstime} but also provides a clear indication that, as may be expected from the dynamical nature of the scattering mechanism, our wires are predominantly in the diffusive (as opposed to localisation) regime.  Indeed, a nonconducting system, such as an insulator or an Anderson localised wire, would be characterised by a vanishing Fermi density of states and therefore, from equation~(\ref{eq_current_density}), a vanishing spatial current variance.  

\begin{figure}[t!]
\includegraphics[trim=0cm 0cm 0cm 0cm, clip=true, width=0.675\textwidth, totalheight=0.345\textheight, angle=0]{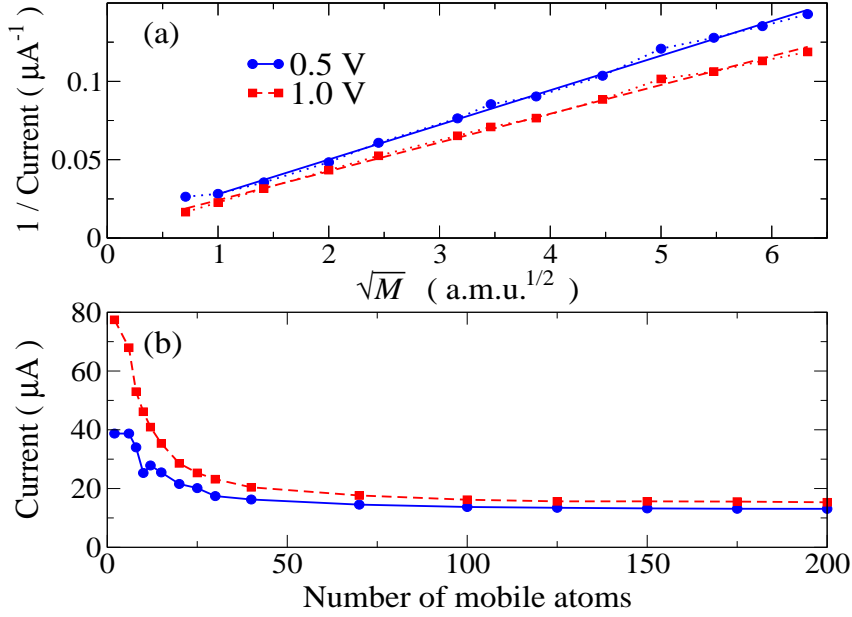}
\caption{(a): Inverse of the average current vs square root of ion mass for a chain with 200 mobile atoms under biases of 0.5~V and 1.0~V.  The straight lines are fits to the data.  (b): Average current for the biases above as a function of $N$ for ions with mass 10 a.m.u.  Notice that, for small $N$, the conductance is almost equal to the quantum unit.}\label{current_image}
\end{figure}

The same general trends as in figure~\ref{kevstime} were observed in all simulations.  Our next task is to investigate the macroscopic characteristics -- namely the total ionic kinetic energy and mean current -- in the long-time saturation regime as a function of wire length, bias and atomic mass.  The defining characteristic of this regime is that the nonconservative forces are counterbalanced by the electronic friction.  The electronic friction is proportional to velocity, $\omega\Delta R$, where $\Delta R$ is a typical ion displacement and $\omega$ is a typical frequency.  The results in section \ref{res_st} indicate that for large enough lengths we can notionally think in terms of a typical length-independent nonconservative current-induced force per atom, roughly proportional to current.  For the kinetic energies in the simulations, vibrational amplitudes are still only a fraction of a bond length, and the nonconservative force should be roughly proportional to these displacements.  Combining these considerations leads to the bias- and length-independent relation
\begin{equation}\label{I_mass}
I\propto\frac{1}{\sqrt{M}},
\end{equation}
where $I$ is the temporally and spatially averaged current in the saturation regime.  Relation (\ref{I_mass}) is verified in figure~\ref{current_image}(a) where $1/I$ is shown as a function of $\sqrt{M}$ for a system with 200 atoms under biases of 0.5~V and 1.0~V: the relation is clearly linear with only a weak bias dependence.

Relation (\ref{I_mass}) can be expressed as $I=g\hbar\omega/\alpha e,$ where $\alpha$ is a dimensionless constant and $g$ is the quantum conductance unit.  Without loss of generality we take $\omega$ to be the Einstein frequency, which for the given tight-binding model is $w_{\rm E}=\sqrt{K_{\rm E}/M}=0.265~{\rm fs}^{-1}$ for 10~a.m.u..  We can then determine $\alpha$ from the gradient in figure~\ref{current_image}(a) and it is found to be 0.94 for the 0.5~V case and 0.79 for the 1~V case.  With these values of $\alpha$, equation~(\ref{I_mass}) predicts length-independent currents of $14.3~\upmu {\rm A}$ and $17.2~\upmu {\rm A}$, in close agreement with the large length limit in figure~\ref{current_image}(b). 

\begin{figure}[t!]
\includegraphics[trim=0cm 0cm 0cm 0cm, clip=true, width=0.65\textwidth,  totalheight=0.35\textheight, angle=0]{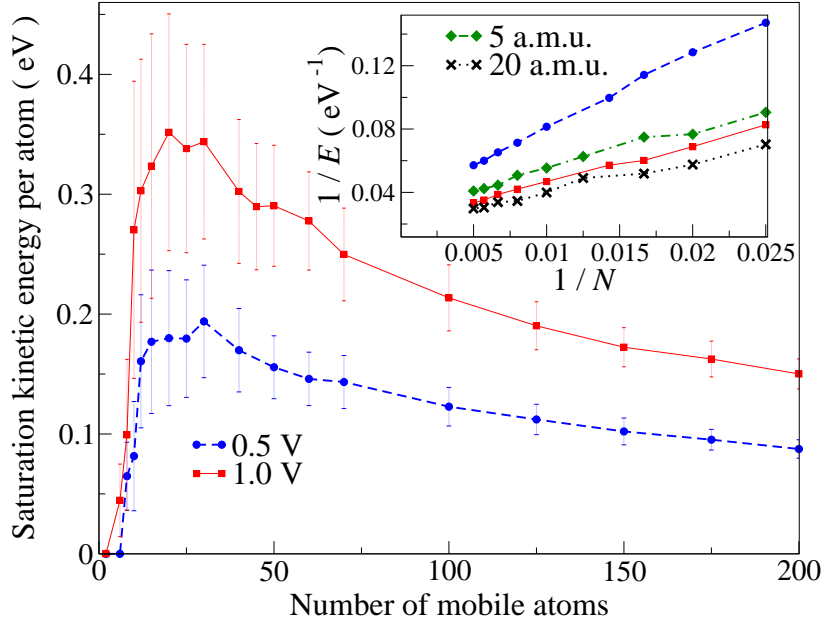}
\caption{Saturation kinetic energy per atom (along with the standard deviation) as a function of the number of mobile atoms in the wire for ions with mass 10~a.m.u., under biases of 0.5~V (blue) and 1.0~V (red).  The inset displays one over the total ionic kinetic energy as a function of $1/N$ for the systems in the main figure, and for masses 5~a.m.u. (green) and 20 a.m.u. (black) under 1~V.}\label{length_image}
\end{figure}

The above considerations are fundamentally a self-consistent condition on the ionic kinetic energy: it must settle at a value producing a resistance such as to make the current agree with relation~(\ref{I_mass}).  Figure~\ref{length_image} shows the saturation kinetic energy per atom as a function of chain length.  We see that this energy, and hence effective temperature, decreases with increasing length for $N$ beyond about 40 atoms.
From the inset, the total kinetic energy has the following dependence on $N$:
\begin{equation}\label{length_eq}
E\approx\frac{E_{N\rightarrow\infty}(V,M)N}{E_{N\rightarrow\infty}(V,M)b(V,M)+N},
\end{equation}
where $E_{N\rightarrow\infty}$ is the bias- and mass-dependent asymptotic value and $b$ is the bias- and weakly mass-dependent slope in the inset.  

These results can be understood as follows.  The current-voltage relation for diffusive conduction in 1-D is \cite{datta:1997,todorov:1996}:
\begin{equation}\label{diffusive}
I=g\frac{V}{1+ L/l},
\end{equation}
where $l$ is of the order of the electronic mean free path for backscattering.  Equation~(\ref{diffusive}) assumes that we are in the linear bias regime (with the present electronic bandwidth of $4|H|=19~$eV, this is likely to be the case under the biases we consider).  Assuming the mean free path to be inversely proportional to the mean square atomic displacements, proportionality between the kinetic and potential energies and approximate equipartitioning of energy between vibrational modes, (\ref{I_mass}) and (\ref{diffusive}) give 
\begin{equation}\label{diff_norm}
\frac{\alpha eV}{1+E/E_0}=\hbar\omega,
\end{equation}
where $E_0$ is a constant.  However, (\ref{diff_norm}) predicts a length-independent total kinetic energy $E$, whereas we already know from (\ref{length_eq}) that this is not the case.  We can make the two agree if we modify (\ref{diff_norm}) to 
\begin{equation}\label{diffusive_image}
\frac{\alpha eV}{1+\displaystyle\frac{E}{E_0}\left(1+\displaystyle\frac{\beta}{N\hbar\omega}\right)}=\hbar\omega,
\end{equation}
with $E_{N\rightarrow\infty}=E_0(\alpha eV/\hbar\omega-1)$ and $b(V,M)=\beta/(\alpha eV-\hbar\omega)E_0$, where $\beta$ is another parameter.  Later we will give an argument to explain the origin of the correction term in brackets in (\ref{diffusive_image}), which will also show that $\beta$ should be of the order of the electron bandwidth.  From the intercepts in the inset in figure~\ref{length_image} we obtain $E_0=16.9$~eV for the 0.5~V, 10~a.m.u. case and $12.9$~eV for the 1~V, 10~a.m.u. case, using the respective fitted values for $\alpha$ above.  The corresponding slopes in the inset give $\beta=23.2~$eV and $18.7~$eV respectively.  According to the model $\alpha$, $E_0$ and $\beta$ should be constants.  The fitted values above show some bias dependence, but it is weak.  Similarly, fitting $E_0$ to the 1~V, 5~a.m.u. and 20~a.m.u. data in figure~\ref{length_image} gives values of 15.5~eV and 9.3~eV respectively, producing a standard deviation of 20\%, for a four-fold variation in mass; the corresponding values of $\beta$, for the green and black data in figure~\ref{length_image}, are 20.8~eV and 12.1~eV, with a similar standard deviation.  \footnote{According to the model in equation~(\ref{diffusive_image}), and in agreement with the inset in figure~\ref{length_image}, the slope should be only weakly dependent on mass at large bias.  The simulation data for the three masses under 1~V in the inset in figure~\ref{length_image} show a small but noticeable scatter, contributing to the variations in the fitted values of the parameters.} Therefore, we regard the fitting as yielding support to the model.

\begin{figure}[t!]
\includegraphics[trim=0cm 0cm 0cm 0cm, clip=true, width=0.69\textwidth,  totalheight=0.36\textheight, angle=0]{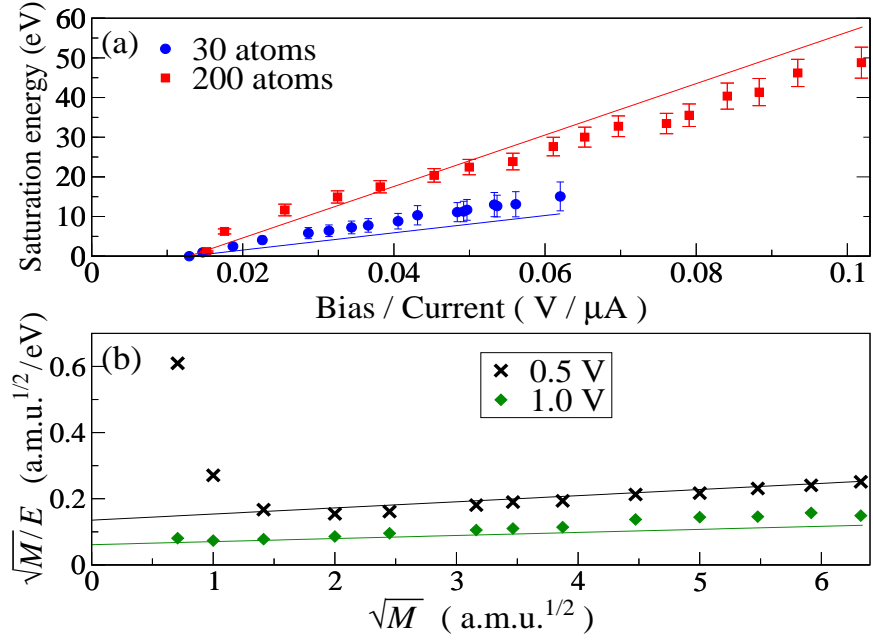}
\caption{(a): Saturation kinetic energy as a function of $V/I$ for biases in the range 0.1 V to 1.7 V in steps of 0.1~V for wires with 30 (blue) and 200 (red) mobile atoms, $M=10~{\rm a.m.u.}$.  The intercept corresponds to $V/I=1/g$, i.e., one quantum unit, in agreement with the critical bias determined from equation (\ref{diffusive_image}).  (b):  $\sqrt{M}/E$ versus $\sqrt{M}$ for $N=200$.  The straight lines are predicted results from relation (\ref{diffusive_image}).  Details are discussed in the text.}\label{bias_image}
\end{figure}

From (\ref{diffusive_image}) we can determine the dependence of energy on applied bias and ion mass.  In figure~\ref{bias_image} we plot the predicted values for the total ionic saturation kinetic energy as a function of $V/I$ in (a) and $\sqrt{M}/E$ versus $\sqrt{M}$ in (b), along with numerical data from the MD simulations.  The predicted results use the parameters from the 1~V 10~a.m.u. case above for the following reasons.  First, 1~V is a representative value for the range of biases in figure~\ref{bias_image}(a).  Second, (\ref{diffusive_image}) can be written as
\begin{equation}\label{trans_eq}
T=\frac{\hbar\omega}{\alpha eV},
\end{equation}
where $T$ is the transmission probability in the saturation regime.  For a given mass, therefore, larger bias takes us further away from the ballistic limit and into the regime for which the above model is designed.  Finally, a mass 10~a.m.u. is representative of the mass range covered in figure~\ref{bias_image}.  The straight lines in figure~\ref{bias_image}(b) are obtained by extracting the linear part of the functional relation between $\sqrt{M}/E$ and $\sqrt{M}$ predicted by equation~(\ref{diffusive_image}) for large $M$. 

In addition to the overall agreement between the model and the simulations, figure~\ref{bias_image} illustrates an important and subtle aspect of the problem.  Since $T$ cannot exceed 1, equation~(\ref{trans_eq}) tells us that for a given mass there should be a critical bias $\sim\hbar\omega/e\alpha$ for nonconservative dynamics to kick in.  This critical bias, furthermore, should correspond to one quantum unit of conductance.  The simulations in figure~\ref{bias_image}(a) clearly show the presence of the critical bias \footnote{This critical bias is to be distinguished from that in figure~{\ref{sum_vs_bias}}, which is needed for the static dynamical response matrix to develop complex eigenvalues.  By contrast, the critical bias in figure \ref{bias_image}(a) is needed for the nonconservative forces to overcome the friction.}.  The ratio of bias to current agrees quantitatively and the value of the critical bias qualitatively.  Conversely, (\ref{trans_eq}) gives a critical mass, $M_{\rm c}\sim K_{\rm E}(\hbar/\alpha eV)^2$, for a given bias, such that for lighter atoms nonconservative effects are suppressed.  This critical mass is the origin of the divergence in the simulation results at small $M$ in figure~\ref{bias_image}(b), although the actual value of $M_{\rm c}$ agrees only to within an order of magnitude.  One can expect this critical region to be difficult to capture in quantitative detail.  Qualitatively though, the presence of a critical bias and a critical mass provides direct criteria for stability against nonconservative dynamics.
\subsection{Further discussion}
We conclude this section with two further arguments to gain additional insight into the problem.  First we consider the correction term in brackets in equation~(\ref{diffusive_image}).  We attribute this correction to small residual localisation effects.  There are different model arguments that lead to the need for this correction.  One {\it ad hoc} argument, which however produces an explicit expression for the correction, is as follows.  First, we can write the transmission function for a 1-D disordered conductor as $T=1/(1+1/\tau)$, where $1/g\tau$ can be interpreted as the resistance of the disordered segment itself \cite{todorov:1996}, and $g\tau$ as its conductance.  For metallic conduction $\tau$ is determined by the conductivity and the system length giving $g\tau=e^2vl_{\rm tr}d/L$.  Next we observe that $l_{\rm tr}d$ is proportional to the number of states, $\Delta N$, available to conducting particles per mean free path.  Next we consider our given transport problem.  The conducting mechanism in operation is likely to be a complicated mix of normal diffusion, with a mean free path $l$, and vibrationally assisted hopping between quasi-localised states, of typical spatial extent $l$.  The net effect, however, is that the motion of the electrons can be thought of as a random walk of typical hopping length $\sim l$.  Based on the earlier current-noise analysis, and on the simulation results, we assume this mechanism remains sufficiently close to metallic conduction, to enable the above characterisation in terms of the quantity $\Delta N$.  Even though vibrations are classical, it remains true that electrons exchange energy with the vibrations in amounts of $\pm \hbar\omega$ (through stimulated emission/absorption) per scattering event.  Here, as before, $\omega$ is a typical vibrational angular frequency. Consider an electron that has made it to the bulk of the conductor, close to the middle.  This has required of the order $n$ hops, where $\sqrt{n}\sim L/2l$.  As a result of the energy exchange with vibrations, its energy would have drifted through a root mean square amount $\Delta E\sim \sqrt{n}\hbar\omega\sim L\hbar\omega/2l$.  The typical energy separation between quasi-localised levels, within a segment $\sim l$, will be of the order of $\beta/(l/R)$, where $\beta$ is the electronic bandwidth.  Therefore our electrons in the bulk can access $\Delta N\sim L\hbar\omega/2\beta R$ of those states per hop.  This reasoning assumes that $\Delta E$ is of the order of, or larger than, the energy spacing $\beta/(l/R)$, and that therefore $\Delta N$ should not be much less than unity. This places limits on how small $L$ or $\omega$ can be for this argument to apply. Notice that $\Delta N$ is independent of $l$ (and therefore of the details of the diffusion mechanism), although $\Delta E$ is not.
Next, consider normal diffusion, that is, diffusion without localisation corrections.  The same number of states, $\Delta N\sim L\hbar\omega/2\beta R$, just found above remain available to diffusing particles, due to the energy exchange with phonons.  In addition, however, further states will be available that were not present, or at least were suppressed, above.  These are the states that the normal metallic density of states provides to Fermi electrons, in the localisation-free conductor.  If $l/v$ is the hopping time then the number of these additional states, accessible per mean free path, will be $\sim(\hbar/(l/v))(l_{\rm tr}/R)/\beta$, where we have made use of the uncertainty principle.  \footnote{Possibly a better picture is the converse. Quasi-localised states that are nearby spatially tend to avoid each other in energy, and vice versa. Thus, it is not so much that the metallic case has an excess of states, but rather that the quasi-localised case has a {\it relative} deficit, in the surrounding {\it local density of states}, as a result of the correlation between position and the energies of nearby states. Either way, it is this relative difference that we are describing.} In our case $v\sim\beta R/2\hbar$, giving a constant number $\eta$, of order unity, of these additional states.  Now, therefore $\Delta N\sim\eta+L\hbar\omega/2\beta R$.  Since $\Delta E\sim[L\hbar\omega/2l]$ is the same in each case, the ratio of disordered-segment conductances, in the normal and the present, partially thermally assisted case, is $\sim(1+2\eta\beta/N\hbar\omega)$, where $N=L/R$.  Finally in the normal case we write $1/\tau=L/l$, where $l$ is the ordinary diffusional mean free path for backscattering.  For the present case, this then gives
\begin{equation}
T=\frac{1}{1+\displaystyle\frac{L}{l}\left(1+\frac{2\beta\eta}{N\hbar\omega}\right)},
\end{equation}
which, in essence, is our desired result, with the additional insight that the parameter $\beta$ considered in the fitting earlier should be of the order of the bandwidth.  Indeed, the fitted values for that parameter are close to the bandwidth, $4|H|$.

Finally we want to make a connection with the parameter $E_0$ introduced earlier.  To this end, we make a standard estimate of $l$.  A simple, but physical, representation of the vibrations, from the point of view of the electrons, is to treat each bond as an independent oscillator.  Then a straightforward Fermi golden rule calculation gives
\begin{equation}
\frac{R}{l}\approx \displaystyle\frac{H'^2}{H^2\sin^2{(\nu\pi)}}\langle X^2\rangle,
\end{equation}
where $\langle X^2\rangle$ is the mean square variation in bond length and $\nu$ is the band filling.  Let our bonds have an effective stiffness $K_{\rm eff}$.  Assuming equipartioning between potential and kinetic energy, the total vibrational kinetic energy in the system is $NK_{\rm eff}\langle X^2\rangle/2$.  Then $L/l=E/E_0$, where
\begin{equation}
E_0=K_{\rm eff}H^2\sin^2{(\nu\pi)}/2H'^2.
\end{equation}
For our tight-binding parameters, this gives $E_0=R^2K_{\rm eff}\sin^2{(\nu\pi)}/2q^2$.  Setting this equal to the fitted value for $E_0$ above gives $K_{\rm eff}\sim 88.4~{\rm eV\AA^{-2}}$.  If we substitute this effective bond stiffness into a nearest neighbour spring model, we obtain a phonon bandwidth of $\sqrt{4K_{\rm eff}/M}\sim0.58~{\rm fs}^{-1}$ for a mass of 10~a.m.u., in reasonable agreement with the actual bandwidth seen in figure~\ref{eqm_freq}(a) (of the order of 0.4~${\rm fs}^{-1}$).  This model argument tells us what factors contribute to $E_0$.  An improved estimate would have to take account of the actual phonon band structure, together with the fact that Fermi electrons typically interact with phonons with a particular wave vector ($\sim2k_F$, where $k_F$ is the Fermi wavevector of the electrons).  In addition to the equipartitioning between potential and kinetic energy, the above argument implicitly assumes equipartitioning of energy between different vibrational modes, which may or may not be obeyed under the nonconservative forces.  This is an interesting avenue for further research.
%
%
\section{Summary}

This study demonstrates that defect-free metallic nanowires are a promising test-bed for nonconservative current-driven dynamics on a grand scale.  We have seen that this is an intricate problem from a physical point of view.  But, in addition, these effects raise the question of stability.  The above findings furnish practical criteria for the likely regions of stability.  Increasing wire length reduces the saturation energy per atom, as does decreasing mass.  The critical bias and mass, below which the nonconservative effect is suppressed, define a transition between dramatically different regimes.  

There are numerous interesting directions for further work.  First, as explained earlier, the present simulations exclude a key physical process: Joule heating.  The interplay between Joule heating and nonconservative forces is an exciting problem.  In the present case, however, in the saturation regime Joule heating should not change the dynamics appreciably.  The reason is that Joule heating results from spontaneous phonon emission; the nonequilibrium contribution to which should scale as $(e\widetilde{V}-\hbar\omega)$, where $e\widetilde{V}$ is the effective scaled bias, $\hbar\omega/\alpha$ (corresponding to the saturation current, as seen from equation~(\ref{trans_eq})).  Since $\alpha\approx 1$, the spontaneous phonon emission rates should be small in the regime considered.  

It is tempting to consider what happens in the limit where the correction term in equation~(\ref{diffusive_image}) is very large.  However, as explained above, our present argumentation does not allow us to venture in to that limit.  

The non-steady-state current fluctuations are a curious phenomenon where, however, electron-electron screening is likely to play a central role.  It would tend to screen out the driving fields due to vibrations and suppress charge fluctuations and hence the nonadiabatic current fluctuations.  Another direction is the Peierls instability that tends to occur under compression-free conditions and ensuing dynamics in the presence of the resultant band gap.  We hope that the present work will motivate further research into some of these questions. 

\section{Acknowledgements}
We are grateful for support from the Engineering and Physical Sciences Research Council, under grant EP/I00713X/1.  This work used the ARCHER UK National Supercomputing Service (http://www.archer.ac.uk).
\bibliography{references}
\end{document}